\documentstyle[12pt]{article}
\def\1ad{\mbox{\normalsize $^1$}}
\def\2ad{\mbox{\normalsize $^2$}}
\def\3ad{\mbox{\normalsize $^3$}}
\def\4ad{\mbox{\normalsize $^4$}}
\def\5ad{\mbox{\normalsize $^5$}}
\def\6ad{\mbox{\normalsize $^6$}}
\def\7ad{\mbox{\normalsize $^7$}}
\def\8ad{\mbox{\normalsize $^8$}}
\def\makefront{\vspace*{1cm}\begin{center}
\def\newtitleline{\\ \vskip 5pt}
{\Large\bf\titleline}\\
\vskip 1truecm
{\large\bf\authors}\\
\vskip 5truemm
\addresses
\end{center}
\vskip 1truecm
{\bf Abstract:}
\abstracttext
\vskip 1truecm}
\newcommand{\be}[3]{\begin{equation}  \label{#1#2#3}}     
\newcommand{\bib}[3]{\bibitem{#1#2#3}}
\newcommand{\ee}{ \end{equation}}
\newcommand{\ba}{\begin{array}}
\newcommand{\ea}{\end{array}}
\newcommand{\p}{\partial}
\newcommand{\NP}[3]{{\em Nucl. Phys.}{ \bf B#1#2#3}}

\newcommand{\PL}[3]{{\em Phys. Lett.}{ \bf B#1#2#3}}

\begin {document}

\rightline{HUB-EP-98/03}
\rightline{hep-th/9801058}

\def\titleline{
Branes in $N=2$, $D=4$ supergravity and the
\newtitleline
conformal field theory limit\footnote{Based on
a talk given at the ``31st International Symposium Ahrenshoop
on the Theory of Elementary Particles'' 
Buckow, September 2-6, 1997.}
}
\def\authors{
Klaus Behrndt
}
\def\addresses{
Institut f\"ur Physik, Humboldt-Universit\"at, \\
Invalidenstr. 110, D-10115 Berlin
}
\def\abstracttext{
In this article we summarise the brane solutions (instantons, black
holes and strings) in 4 dimensions when embedded in $N=2$
supergravity. Like in 10 dimensions these solutions are related by
duality transformations ($T$-duality, $c$-map). For the case that the
graviphoton has no magnetic charge, we discuss the conformal field
theory description of the near-horizon geometry. The
decoupling of the massless modes in the matrix limit of
the $M$-theory compactified on a Calabi-Yau threefold
is discussed.
}
\makefront

In the last years we made substantial progress in the understanding of
$N$=2 black holes in 4 and 5 dimensions \cite{010,020,070,080,090}. In order
to obtain these black holes one has to compactify the 10-dimensional
supergravity on an internal space with unknown metric, $K3$ for heterotic
and Calabi-Yau threefold ($CY_3$) on type II side. As consequence, one
cannot obtain them simply by dimensional reduction and hence one has to
solve either the 4-d equations of motion or the Killing spinor equations.

Considering only the ungauged bosonic part the equations of motion follow
from the $N$=2 supergravity action given by
\be010
S = \int d^4 x \, \sqrt{G} \, \left\{
R - \frac{1}{4} F^{I}_{\mu\nu}\,^{\star} G_{I}^{\ \mu\nu}
- 2 g_{A \bar{B}} \p_{\mu} z^A \p^{\mu} \bar z^B
- \tilde{g}_{uv} \p_{\mu} q^u \p^{\mu} q^v \right\}\,.
\ee
The $n_v$ complex scalars $z^A$ coming from vector multiplets
parameterize a special K\"ahler manifold $K_{n_v}$ with metric $g_{A
\bar{B}}(z, \bar{z})$, whereas the $4n_h$ real scalars $q^u$
a quaternionic manifold $Q_{n_h}$ with metric
$\tilde{g}_{uv}(q)$.  The gauge field strengths ($F^I_{\mu \nu} ,
G_{I\, \mu\nu}$) form a symplectic vector ($I= 0 ... n_v$) and
therefore are not independent
\be020
G_{I\, \mu\nu} = \mbox{Re} {\cal N}_{IJ} F^{J}_{\mu\nu} - 
 \mbox{Im} {\cal N}_{IJ} {^{\star}}F^J_{\mu\nu} \ .
\ee
Like in 10 dimensions we can classify the 4-d supergravity theory in
terms brane-solutions (related to elementary and solitonic states); for
the 6-d case an analog discussion can be found in \cite{040}. These are
the membranes, strings, black holes (0-branes) and instantons
(-1-branes). The membranes introduce a mass term and are the analog
of the 8-brane solution of massive IIA supergravity.  We will ignore this
solution furthermore and focus mainly on the other three branes.  
The string solution is naturally supported by antisymmetric tensors, that
couples to their worldvolume. In the action (\ref{010})
the antisymmetric tensors have been dualized to scalars which are
part of the hyper scalars. Black holes couple naturally to vector fields
and therefore are solutions with non-trivial vector
multiplets. Finally the instantons are dual to the strings and
are related to non-trivial hyper scalars.  In addition to these
counterparts of the 10-d $D$-branes, we have also purely gravitational
solutions, the $G$-branes as discussed in \cite{050}, the wave and
Taub-NUT soliton. The branes can also be seen as a classification of
the solutions in the number of translational isometries.

These are the ``basic'' branes of 4-d supergravity.  In addition, 
(threshold) bound states should exist, but so far no solution could be
constructed which includes non-trivial vector as well as hyper multiplet
scalars. But we can adopt the 10-d duality rules to transform all
``basic'' solutions into one another.

A good starting point is the black hole or 0-brane. In this case we can
neglect the quaternionic part by taking $q^u = constant$. Let us
summarise here only the main points, for details we refer to \cite{070}.
The gauge field equations and Bianchi identities are given by $ dF^I =
dG_I = 0$ and have the solution
\be040 
F^I_{mn} = {1 \over 2} \epsilon_{mnp} \partial_p \tilde{H}^I
\quad , \quad G_{I\, mn} = {1 \over 2} \epsilon_{mnp} \partial_p H_I
\ee
where $m,n,p= 1,2,3$ and $(\tilde{H}^I , H_I)$ are harmonic functions.
Note, that the timelike components $F^I_{0m}$ and $G_{I\, 0m}$ are
fixed in terms of the spatial ones using (\ref{020}). As next step
we need the scalars and the metric, which can be obtained by solving
the Killing spinor equations \cite{070}
\be070
ds^2 = - e^{2U} (dt + \omega_m dx^m)^2 + e^{-2 U} dx^m dx^m 
\qquad , \qquad z^A = {X^A \over X^0}
\ee
with
\be080
\ba{rcccl}
e^{-2U}&=&e^{-K} & \equiv & i (\bar{X}^I F_I- X^I \bar{F}_I) \ , \\
{1 \over 2} e^{2U} \epsilon_{mnp} \partial_n \omega_p&=&Q_m & \equiv &
{ 1 \over 2} e^{2U} ( H_I \partial_m \tilde{H}^I - \tilde{H}^I
\partial_m H_I)
\ea
\ee
and the symplectic section is constraint by
\be090
-i (X^I - \bar{X}^I ) = \tilde{H}^I(x^{\mu}) \qquad , \qquad 
-i (F_I -\bar{F}_I) = H_I(x^{\mu}) 
\ee
where the harmonic functions $(\tilde H^I , H_I)$ are introduced in
(\ref{040}) and $F_I = {\p \over \p X^I} F(X)$. This solution is
expressed completely in terms of duality invariant quantities, the
K\"ahler potential $K$ and the K\"ahler connection $Q_m$. But it is {\em
not} written in a K\"ahler invariant way, instead we have fixed the
K\"ahler invariance in a proper way. If $\omega_m = 0$ we get the
expected static black hole and for special choices for the harmonic
functions we obtain a rotating black hole or generalisations of Taub-NUT
and Eguchi-Hanson spaces.

By $T$-dualizing different directions of the 4-d space time, one obtains
the other brane solutions. In order to be concrete let us consider one
simple example, which is axion-free, i.e.\ ${\rm Re} z^A = 0$.
This can be done by setting $H^0 = H_A =0$, i.e.\ the graviphoton is
electric and all vector multiplets are magnetic. As prepotential we
consider
\be100
F(X) = { d_{ABC}X^A X^B X^C \over X^0}
\ee
and the solution is given by \cite{080}
\be110
e^{-2U} = \sqrt{H_0 \, d_{ABC} H^A H^B H^C} \qquad , \qquad 
z^A = i\, H_0 H^A e^{2U}
\ee
with $H_0 = h_0 + q_0/r$ , $H^A = h^A + p^A/r$. This black hole is a
compactification of the $5 \times 5 \times 5 + mom.$ intersection, where
each of the 5-brane wraps a 4-cycle of the $CY_3$ and $H_0$ corresponds to
the momentum modes travelling along the common string. A detailed
microscopic analysis has been given recently in \cite{020}.

In order to obtain the instanton solution we have first to $T$-dualize
the Euclidean time, which is also known as $c$-map has been used in
\cite{090} to map the general solution (\ref{080}).  Note, because the
black hole solution considered here, is a type IIA compactification,
after this mapping we obtain a type IIB compactification, where the
wrapped 5-branes become (instantonic) wrapped 3-branes and the
momentum modes are converted into a IIB (-1)-brane, i.e.\ the
intersection is $(-1)\times 3 \times 3 \times 3$.
The resulting solution reads \cite{090}
\be120
\ba{l}
ds^2_{inst.} = e^{-2U} dx^{\mu} dx^{\mu} \quad , \quad
e^{2 \phi} = e^{-2U} \quad , \quad \zeta^0 = {1 \over \sqrt{2}} A^0 = { 1 \over
\sqrt{2} H_0} \\  \p_m \zeta_A = - { 1 \over 2 \sqrt{2}} e^{2U} {\rm Im}
{\cal N}_{AB} \p_m H^A \qquad , \qquad z^A \rightarrow q^A 
\ea
\ee
where the first line contains the string-metric and the universal
hypermultiplet, i.e.\ the dilaton $\phi$ and the former graviphoton gauge
potential. In the second line are the scalars coming from the dualization
of the magnetic vector fields and also the original vector multiplet
scalars. All these scalars group together in $n_v +1$ hypermultiplets
each containing 4 scalars.  After having done this dualization one can
allow the fields to depend on all 4 coordinates in a symmetric way, i.e.\
the new harmonic functions are $H_0 = h_0 + q_0/r^2$ , $H^A = h^A +
p^A/r^2$. Note, the Einstein metric is flat.

As next step we consider the string solution, obtained by
$T$-dualizing one of the spatial direction. This is again a type IIB
solution corresponding intersection is $1 \times 5 \times 5
\times 5$
\be130
\ba{c}
ds^2_{string} = e^{2U} \left( -dt^2 + dx^2 \right) + e^{-2U} dw d\bar w \\
e^{2 \phi} = e^{2U} \  , \  \p_m a \sim \epsilon_{mnp} \p_p H_0 \  ,
\  {\cal H}_A \sim e^{2U}{\rm Im}{\cal N}_{AB}\,^{\star} d H^B 
\  , \  \tilde {\cal H}_A \sim -i\,e^{2U} g_{AB}\,^{\star} dz^B
\ea
\ee
where we choose a parameterization in terms of one complex scalar
containing the axion $a$ and the dilaton $\phi$. In addition, the
compactified $D$-5-branes and the Hodge dualization of the scalars give
$2n_v$ antisymmetric tensors $({\cal H_A} , \tilde {\cal H}_A)$. Equivalently,
one could dualize them into (hyper) scalars. In doing this $T$-duality we
have to assume a further isometry, i.e.\ the harmonic functions depend on
one coordinate less. Thus the new harmonic functions are general
holomorphic functions $H = f(w) + f(\bar w)$, i.e.\ they are undetermined
from the equations of motion, but restricted due to the proper
behaviour at infinity and finite energy, see \cite{100}. Notice however,
because of the K\"ahler structure it is not obvious that the dilaton
factorizes in harmonic factors. We will leave a detailed analysis for the
future.

One has to keep in mind, that for all these solutions the internal
space is fixed (branes wrapped around 4-cycles). Our $T$-duality
transformation acts in the 4-d space time. Duality with respect to the
internal manifold would of course not change the 4-d solution, it is
given by the duality invariant K\"ahler potential. 

Having these ``basic'' brane solutions we have to ask for singularities.
Let us focus on the instantons and black holes. The harmonic function
that we consider here have a pole of $1/r$. As consequence of the
constraint for the symplectic section (\ref{090}) also $(X^I , F_I)$ have
a $1/r$ pole. Thus the K\"ahler potential behaves like $1/r^2$ and the
metric is non-singular on the horizon. This property is independent of
the number of charges, as long as we keep a non-trivial graviphoton. So,
{\em any generic CY black hole has a non-singular horizon}. This is an
important difference to $N=4,8$ black holes that need at least 4 charges
to get a non-singular horizon. Here, in the CY case, the selfintersection
regularize the horizon, which has a throat geometry.
Of complete different nature is the instanton solution.
Again using the scaling behaviour of the K\"ahler potential one can show,
that for the $r \rightarrow 0$ one reaches a second asymptotic flat
region, where the charges and moduli are interchanged. Both regions are
connected by wormhole at the minimum of $re^{-U}$ and one of them is
strongly coupled $(\phi \rightarrow \infty)$, which is very similar
analog to the situation in 10 dimensions \cite{030}.

Let us end this article with the conformal field theory (CFT)
description of the near horizon region of the black hole. Basically it
is not necessary to approach the horizon, instead one can scale the
physical parameter approprietely (conformal limit). The result is the
same -- one neglects the constant parts in the harmonic functions in
this limit. All non-singular brane configurations are expected to have
a super conformal field theory description \cite{160}. As example we
consider the solution (\ref{110}).  This is a special case, but if we
convince ourselfes that it is described by an exact CFT, due to
duality also the other charge configurations should be exact. This
solution is a compactified 5-d magnetic string \cite{080}.  Near the
horizon, the spherical part becomes constant and a reduction over this
spherical part yields a (negative) cosmological constant giving an
$AdS_3$ geometry. By proper boundary conditions a conformal algebra is
realized as diffeomorphism group at spatial infinity \cite{130} (this
is also nicely summarized in \cite{110}) and the corresponding solution
is the BTZ black hole \cite{120} (for a discussion of the entropy and
duality see especially \cite{150,110,140}).

In the following we will sketch this procedure for our solution, a
more detailed analysis will be given elsewhere.  Near the horizon the
magnetic string becomes \cite{080}
\be140
ds^2 = {r \over D^{1/3}} \left[ -dt^2 + dx^2 + {q_0 \over r}
 (dx -dt)^2 \right] + D^{2/3} \left({dr \over r}\right)^2 + 
 D^{2/3} d\Omega_2 \ ,
\ee
where $D= d_{ABC}p^A p^B p^C$. As next step one compactifies the $S^2$
part. The 5-d sugra action contains besides the Einstein-Hilbert term, a
gauge field part and a scalar part. Near the horizon the scalar part
drops out and the gauge field part contributes to the cosmological
constant $G_{IJ}F^I \cdot F^J = D^{-2/3}$ with $G_{IJ}(X)$ as the 5-d gauge
coupling matrix. After this reduction the 3-d Einstein metric 
reads
\be160
ds^2_E = D\, r \left[ -dt^2 + dx^2 + {q_0 \over r}
 (dx -dt)^2 \right] + D^2 \left({dr \over r}\right)^2 \ 
\ee
or in other coordinates ($\rho^2 = D (r + q_0)$, $\tau = 2 D t$ and
$x = \phi$)
\be170
ds^2_E = - {(\rho^2 - q_0 D)^2 \over 4 \rho^2 D^2} d\tau^2 + \rho^2
\left(d \phi - {q_0 \over 2 \rho^2} d\tau \right)^2 + { 4 D^2 \rho^2 \over 
(\rho^2 - q_0 D)^2} d\rho^2 \ .
\ee
This is the extreme 3-d BTZ black hole \cite{120}, with the angular
momentum $J = q_0$, the mass $M = {q_0 \over 2 D}$ and the cosmological
constant $\Lambda^{-1} \sim l = 2D$. Thus excited momentum modes along
the 5-d string correspond to angular momentum for this BTZ black hole. As
it has been shown in \cite{200}, 3-d AdS gravity can be written as a
Chern-Simons theory and furthermore, due to the cylindric symmetry ($\phi
\simeq \phi + 2 \pi$) it reduces to a sum of two SL(2,R) WZW
theories for the left and right moving modes (in the BPS limit only one
sector survive). Finally, implementing the boundary condition at spatial
infinity we end up with a Liouville theory living on the boundary
\cite{210}. So, summarizing all steps we have the sequence for the
actions
\be180
S_{sugra_5} \rightarrow S_{AdS_3} \; \sim \; k \, S_{CS} 
\; \sim \; k \,S_{WZW} \; \sim \; k \, S_L
\ee
where up to numerical factors, the central charge of the CFT is
given by  $k \sim {D \over G_N}$, where we introduced the 3-d Newton
constant $G_N$ to get a dimensionless quantity.

Let us conclude with implications of this CFT for the Matrix
description.  It has been suggested recently that the matrix limit of
$M$-theory compactified on $CY_3$ is given by the conifold point,
which is mirror dual to the vanishing 6-cycle \cite{190}. An important
point in the matrix limit is to ensure the decoupling of massless
states (decoupling from the bulk). For a $T^6$ compactification this
is in general not the case (although as discussed in \cite{180} there
is may be a loophole by a gauged world volume theory). In order to
discuss this question here we consider only 5-branes which wrap
vanishing cycles. Notice, in our setup we approaching the conifold
point from the other side, i.e.\ we consider vanishing 4-cycles.  As
consequence, the corresponding 4-d black hole besomes massless
\cite{230} or, for the magnetic sting, $D \rightarrow 0$ in this limit 
(note, if $L$ counts the vanishing cycles $X^L \sim p^L \rightarrow 0$
at this point). As consequence the central charge of the CFT vanish,
which coincides with the expectation that at the transition point the
worldvolume theory is scale invariant. But more important, in
this limit the boundary degrees of freedom (encoded in the Liouville
action $S_L$) decouple, like the Liouville mode in critical string 
theory. This supports the idea that the matrix description for $M$-theory
compactified on $CY_3$ \cite{190} yields an consistent picture.

If one goes away from the conformal point one has to take into account
the additional boundary modes and an immediate question is to
ask for their contribution to the black hole entropy of (\ref{110}). 
Of course, quantizing the boundary CFT one obtains a shift
in the central charge (from the Liouville mode). This is equivalent to a
shift in the cosmological constant $D \rightarrow D + const.$, or the
corrected black hole entropy reads
\be190
{\cal S} \sim 2 \pi \sqrt{q_0 (d_{ABC}p^A p^B p^C + const.)} \ .
\ee
It would be very interesting to discuss these corrections in more detail.
Note this shift is independent of the magnetic charges that parameterize
the the 4-cycles of the CY. Similar shifts of $D$ have recently been
discussed in terms of higher loop corrections for the black hole entropy
\cite{020,340}, but the corrections there are linear in the
magnetic charges. Because the shift here is not sensitive to any internal
CY structure, one should expect that it is a pure supergravity
effect, i.e.\ related to terms in the prepotential that contain only
$X^0$. A further interesting line of continuation
would be to keep the scalar fields as matter part in this CFT description.

\bigskip

\noindent
{\bf Acknowledgements}
\medskip 

\noindent
I am grateful I.\ Brunner, G. Cardoso, S.\ Kachru,
A. Karch and R. Schimmrigk
for helpful discussions. This work is support by the DFG.


\end{document}